\documentclass[12pt,letterpaper]{article}
\pdfoutput=1

\input{common/preamble}

\usepackage{lineno}

\begin{document}

\pagestyle{titlepage}


 \pagestyle{titlepage}

\date{}

\title{\scshape\Large 
Neutrino Frontier Topical Group Report (NF03): \\[2mm] Physics Beyond the Standard Model
\normalsize 
\vskip -10pt
\snowmasstitle
}


\renewcommand\Authfont{\scshape\small}
\renewcommand\Affilfont{\itshape\footnotesize}
\renewcommand\Authand{, }
\renewcommand\Authands{, }

\author[1]{P.~Coloma}
\author[2]{L.~Koerner}
\author[3]{I.~Shoemaker}
\author[4]{J.~Yu}
\author[ ]{\\on behalf of the NF03 Topical Group Community$^{\thanks{This report is based on the contributed whitepapers ~\cite{Abdullahi:2022jlv,Arguelles:2022xxa,Berryman:2022hds,Abdullah:2022zue,Dev:2022jbf,Berger:2022cab,Batell:2022xau}.}}$ }

\vspace{-0.5cm}
\affil[1]{Instituto de F\'isica Te\'orica UAM-CSIC, Universidad Aut\'onoma de Madrid, Calle de Nicol\'as Cabrera 13–15, Cantoblanco, E-28049 Madrid, Spain}
\affil[2]{University of Houston, Houston, TX 77204, USA}
\affil[3]{Center for Neutrino Physics, Department of Physics, Virginia Tech University, Blacksburg, VA 24601, USA}
\affil[4]{University of Texas at Arlington, Arlington, TX 76019, USA}

\maketitle

\begin{abstract}
    This topical group focuses on searches for signals from physics Beyond the Standard Model (BSM), both from a theoretical and experimental perspective, including neutrino-related BSM and searches for BSM in other sectors using neutrino facilities.
\end{abstract}

\renewcommand{\familydefault}{\sfdefault}
\renewcommand{\thepage}{\roman{page}}
\setcounter{page}{0}

\pagestyle{plain} 
\clearpage
\textsf{\tableofcontents}





\renewcommand{\thepage}{\arabic{page}}
\setcounter{page}{1}

\pagestyle{fancy}

\fancyhead{}
\fancyhead[RO]{\textsf{\footnotesize \thepage}}
\fancyhead[LO]{\textsf{\footnotesize \rightmark}}

\fancyfoot{}
\fancyfoot[RO]{\textsf{\footnotesize Snowmass 2021}}
\fancyfoot[LO]{\textsf{\footnotesize NF03 Topical Group Report}}
\fancypagestyle{plain}{}

\renewcommand{\headrule}{\vspace{-4mm}\color[gray]{0.5}{\rule{\headwidth}{0.5pt}}}



\clearpage

\section*{Executive Summary}
\label{sec:summary}

The arrival of the ``precision era'' in neutrino physics comes with both challenges and opportunities.  One of these opportunities is the ability to search for evidence of physics Beyond the Standard Model (BSM), expanding the physics scope of current and future experiments well beyond the measurement of three-flavor neutrino oscillations. In fact, while the SM describes with great accuracy most of the phenomena observed in high-energy physics, a number of fundamental scientific questions still need to be answered:
\begin{enumerate}
    \item Is the dark sector as complex and phenomenologically rich as the visible sector? How does the dark sector interact with the visible sector? 
    \item What are the properties and interactions of neutrinos, and how many are there? What is the explanation behind the observed anomalies in short-baseline neutrino experiments? 
    \item What is the mechanism responsible of neutrino mass generation? Are lepton and baryon numbers fundamentally conserved, or just accidental symmetries? 
    \item What is the mechanism responsible for the matter antimatter asymmetry of the Universe?
    \item Do all the forces unify? If so, at what energy scale? 
\end{enumerate}

Current and future neutrino experiments are in an excellent position to test a wide variety of BSM models not only in the neutrino sector, but in other sectors, as well. In particular, the high-intensity proton beams and massive precision detectors enable BSM physics searches at future experiments whose primary goals are to measure neutrino oscillation parameters, providing the possibility to expand their physics reach in a way that is complementary to Energy Frontier experiments. This report summarizes those opportunities in BSM physics searches, divided into two main  categories as follows:
\begin{itemize}
\item {\bf Neutrino-related BSM scenarios}
\begin{itemize}
    \item {\bf Heavy Neutral Lepton Searches}\cite{Abdullahi:2022jlv}.  These include searches for the HNL decay products at colliders and extracted beamlines, as well as similar searches using neutrinos from atmospheric, solar, and conventional neutrino beams. Competitive bounds can also be derived from nuclear decay searches, and from the impact of HNL on cosmological observables, among others. The Electron-Ion Collider can also play a key role in probing HNLs~\cite{Batell:2022ubw}.
    
    \item {\bf BSM Effects on Neutrino Flavor}\cite{Arguelles:2022xxa}. Precision measurements of neutrino oscillations, as well as measurements of the flavor composition and spectra of astrophysical neutrinos, allow precision tests of the three-neutrino paradigm. These can be used to derive powerful constraints on a variety of BSM scenarios, including non-unitarity of the leptonic mixing matrix, large extra dimensions, non-standard neutrino interactions and light mediators, long-range forces, neutrino decay, decoherence, CPT and Lorentz violation. Neutrino-dark matter interactions may also lead to visible effects in oscillation experiments. 
    
    \item{ \bf BSM Effects on Neutrino Scattering.}  Neutrino self-interactions~\cite{Berryman:2022hds} via light mediators can lead to observable effects in supernovae, high-energy astrophysical neutrino spectra, and measurements of neutrino scattering. The existence of light mediators coupled to neutrinos can also lead to significant modifications in cosmological observables.  The recent observation of Coherent $\nu$-Nucleus Elastic Scattering (CE$\nu$NS)~\cite{Abdullah:2022zue} has opened new opportunities to search for new physics in the neutrino sector, especially regarding the existence of new neutrino interactions.  
    
\end{itemize}
\item {\bf Non-neutrino BSM scenarios}
\begin{itemize}
    \item {\bf Baryon Number Violation Searches}\cite{Dev:2022jbf}. The use of very massive detectors in neutrino experiments enables strong constraints on proton decay and neutron-antineutron oscillations. Both searches can shed light onto the mechanism behind the matter antimatter asymmetry of the Universe and onto the question of unification of forces. 
  
    \item {\bf Cosmogenic Dark Matter and Exotic Particle Searches}\cite{Berger:2022cab}. A non-minimal dark sector may lead to a wide range of phenomenological observations, including boosted dark matter signals, events from up-scattering and decay of dark sector particles, neutrino lines from dark matter annihilation in the Sun, or scattering signals from slow-moving dark matter. 
    \item {\bf Beam-Originating Dark Sector Particle Searches}\cite{Batell:2022xau}. The use of high-power proton beams may allow production of weakly coupled states together with the neutrino beam. Possible signals include the decays of long-lived-particles into a variety of visible final states, an excess of elastic and/or inelastic scattering with detector electrons or nuclei, or neutrino up-scattering into heavy states followed by their visible decays. 
    \item {\bf Milli-charged particles}\cite{Batell:2022xau}. Milli-charged particles may be produced in the atmosphere or in fixed target experiments and lead to an excess of scattering events in atmospheric detectors or near detectors in conventional beam experiments. 
\end{itemize}
\end{itemize}
A summary of the main phenomenological scenarios described in this document, together with the most relevant experimental signatures for each of them, is provided in Tab.~\ref{tab:my_label}. For reference, we also include a list of some of the planned/proposed experiments where such signals may be investigated (see the provided references for further details). 


Exploring BSM physics in neutrino experiments has traditionally been impractical, if not impossible, in the past. The major challenge that neutrino facilities face in order to be able to set reliable constraints on BSM scenarios concerns the large systematic uncertainties stemming from our imprecise knowledge of both neutrino fluxes and neutrino interaction cross sections. In this sense, the program of supporting measurements in the neutrino frontier (including hadron production measurements for neutrino flux predictions and measurements to improve modeling of neutrino-nucleus interactions) is as important for BSM searches as it is for the standard three-flavor neutrino oscillation measurements. 

In spite of the difficulties mentioned above, neutrino facilities offer a broad range of opportunities for an expansive and strong BSM physics program. Thus, in order to fully exploit their physics reach, BSM searches need to be considered among their main goals since this may impact their design and optimization. 
Future advances in this direction will also require a joint effort between the experimental and theoretical communities. In this sense, data release efforts have empowered the theoretical community in the past to derive new constraints on a wide landscape of BSM models and/or to propose new experimental searches and strategies. In addition, the use of model-independent frameworks and parameterizations to constrain new physics models (a common example is the use of effective operators at low energies, an approach commonly used in other areas of particle physics as well) allows to recast different experimental constraints easily. 

It, however, is equally important that these frameworks are also matched onto viable models, so that the bounds from neutrino experiments can be contrasted with those obtained in other Frontiers of particle physics. A closer collaboration across particle physics frontiers would also be desirable, as many BSM searches at neutrino facilities are complementary to BSM searches in other areas. Moreover, improved simulation tools will be needed to compute the expected BSM signals and backgrounds for which a closer collaboration with the nuclear physics community would be required. 


Finally it is worthwhile emphasizing here that, while the Snowmass 2013 neutrino working group report~\cite{IntensityFrontierNeutrinoWorkingGroup:2013sdv} did include a section on BSM opportunities, it was largely focused on resolving the short-baseline anomalies through the addition of an eV-scale light sterile neutrino. Since then, the community has made a significant effort in pushing the boundaries and the potential of neutrino experiments to test for BSM models widely, and expanding searches to non-minimal BSM scenarios in the neutrino sector. For example, several extensions beyond a standard eV-sterile neutrino have been put forward in the past few years, which are able to reconcile some of the short-baseline anomalies~\cite{Acero:2022wqg} and may lead to additional signals elsewhere. At the same time, the community is heavily exploring the potential of neutrino facilities to test for new physics outside the neutrino sector, which often comes as a result of a close interaction between theorists and experimentalists, as demonstrated in a Ref.~\cite{Arg_elles_2020}. Just to give a couple of examples of this effort in the past few years: strong constraints on non-minimal dark matter scenarios have been set using MiniBooNE data taken in the so-called beam-dump mode~\cite{MiniBooNEDM:2018cxm}, improving over bounds from direct detection experiments in the light mass regime; and new bounds on milli-charged particles have been set by the ArgoNeuT experiment~\cite{ArgoNeuT:2019ckq}.

The present report is the first of its kind, which summarizes a wide range of opportunities to search for BSM physics using neutrino facilities, including dark sector particles produced in conventional neutrino beams or the Sun, heavy neutral leptons with masses at or below a few GeV, and new neutrino properties and interactions, among others.

\afterpage{%
    \clearpage
    \thispagestyle{empty}
    \begin{landscape}
    \centering 
    \begin{table}
    \captionsetup{font=ninept}
    \resizebox{1.18\textwidth}{!}{\begin{tabular}{|l|l|l|l|}
    \hline
    BSM Scenario &  Sources & Signatures  & Example Experiments 
    \\
    \hline 
    \hline
    \multirow{5}{*}{HNL\cite{Abdullahi:2022jlv}} 
    & Colliders  & HNL decay &  ATLAS, CMS, FASER, Belle II, ...  \cr  
    & Nuclear decays  & Nuclear decay kinematics & KATRIN/TRISTAN, HUNTER ...  \cr 
    & Fixed target & HNL decay & DUNE ND, SHiP, ICARUS, ... \cr \cline{2-4} 
    & \multirow{2}{*}{Atm. \& solar $\nu$s} & Distorted recoil spectrum & \multirow{2}{*}{DUNE, HK, IceCube/DeepCore, ...} \cr  
    & & HNL decay, double bangs & \cr \cline{2-4}
    & Early Universe  & Cosmological parameters ($N_{\rm eff}$) & Simons Observatory, CMB-S4, ... 
    \\
    \hline
    Non-unitarity\cite{Arguelles:2022xxa} &  Beam \& Atm. $\nu$s  & Deviations from 3-$\nu$ mixing (ND \& FD) & DUNE, ESS$\nu$SB, HK, ... \\
    \hline
    \multirow{3}{*}{LED\cite{Arguelles:2022xxa}} & 
    Reactor $\nu$s & \multirow{2}{*}{Distortion of oscillated spectra (FD \& ND)} & JUNO, TAO,...\cr 
    & Beam $\nu$s & & DUNE, ... \cr \cline{2-4}
    & Atm. $\nu$s & Anomalous matter effects & Icecube, KM3NeT, ... 
    \\
    \hline
     & 
    Reactor \& Spallation sources & Distortion of CE$\nu$NS rate & COHERENT, CONNIE, CONUS, ... \cr 
    NSI \& light & Solar, Beam, Atm \& SN $\nu$s & Anomalous matter effects & DARWIN, DUNE, T2HKK, HK, IceCube, ...\cr 
     mediators\cite{Arguelles:2022xxa,Abdullah:2022zue} & Beam $\nu$s & Anomalous appearance, $\nu-e^-$ scattering, tridents & DUNE ND, T2HK ND, IsoDAR, ... \cr
    & Collider $\nu$s & Distortion of CC spectrum & FASER$\nu$, ... 
    \\
    \hline 
    Long-range & Solar \& Atm $\nu$ & Anomalous matter potential & HK, JUNO, DUNE, ... \cr 
    forces~\cite{Arguelles:2022xxa} & UHE Astrophysical nus & Distorted flavor ratios & HE Neutrino Telescopes 
    \\ 
    \hline
    \multirow{3}{*}{$\nu$-DM interact. \cite{Arguelles:2022xxa}} & Reactor \& solar $\nu$s & Distorted oscillated spectra, or & JUNO, ... \cr 
    & Beam $\nu$ & time-dependent oscillation params. & DUNE, ... \cr 
    & UHE Astrophysical $\nu$s & Distorted flavor ratios \& spectra & HE \& UHE Neutrino Telescopes
    \\
    \hline
     & SN $\nu$s & SN extra energy loss, distortion in neutrino spectra & DUNE, HK, JUNO, ... \cr
    $\nu$ self & UHE Astrophysical $\nu$s & {Distorted spectra} & HE \& UHE $\nu$ telescopes \cr
    interact.~\cite{Berryman:2022hds, Feng:2022inv} & Early Universe & Effects on CMB, BBN, \& structure formation & CORE, PICO, CMB-S4 \cr 
    & Beam \& Collider $\nu$s & Missing energy \& $p_T$ in $\nu$ scattering & DUNE ND, Forward Physics Facility, ...
    \\
    \hline
    \multirow{4}{*}{$\nu$ decay\cite{Arguelles:2022xxa}} & Reactor \& DAR $\nu$s & \multirow{3}{*}{Distortion of oscillated spectra} & JUNO, IsoDAR, ... \cr 
    & Beam $\nu$s  & & DUNE, MOMENT, ESS$\nu$SB, HK, ...\cr 
    & Atm. $\nu$s  & & INO-ICAL, KM3NeT-ORCA, ... \cr \cline{2-4}
    & UHE Astrophysical $\nu$s & Distorted flavor ratios \& spectra & HE \& UHE Neutrino Telescopes 
    \\
    \hline
    \multirow{3}{*}{CPT violation\cite{Arguelles:2022xxa}} & Beam $\nu$s & \multirow{2}{*}{Different $\nu$ and $\bar{\nu}$ osc. params.} & DUNE, ESS$\nu$SB, HK, ... \cr 
    & Atm. $\nu$s & & IceCube, DUNE, ... \cr \cline{2-4}
    & UHE Astrophysical $\nu$s & Distorted flavor ratios \& spectra & HE \& UHE Neutrino Telescopes  
    \\
    \hline
    \multirow{3}{*}{Lorentz violation\cite{Arguelles:2022xxa}} & Beam $\nu$s & \multirow{2}{*}{Sidereal modulation of event rate} & DUNE, ESS$\nu$SB, HK, ... \cr 
    & Atm. $\nu$s & & IceCube, DUNE, ...  \cr \cline{2-4}
    & UHE Astrophysical $\nu$s & Distorted flavor ratios \& spectra, velocity dispersion & HE \& UHE Neutrino Telescopes 
    \\
    \hline
    \multirow{4}{*}{Quantum decoh.~\cite{Arguelles:2022xxa}} & 
    Reactor \& DAR $\nu$s & 
    \multirow{3}{*}{Distortion of oscillated spectra} & JUNO, IsoDAR, ... \cr 
    & Beam $\nu$s & &  DUNE, ... \cr
    & Atm. $\nu$s & & KM3NeT, IceCube, HK, ...  \cr \cline{2-4}
    & UHE Astrophysical $\nu$s & Distorted flavor ratios & HE Neutrino Telescopes 
    \\
    \hline
    $B$ violation\cite{Dev:2022jbf} & Detector mass & Nucleon decay, $n-\bar{n}$ oscillations & DUNE, HK, JUNO, ...
    \\
    \hline
    \multirow{4}{*}{Dark Matter\cite{Batell:2022xau,Berger:2022cab}} 
    &  DM annihilation, DM decay & Excess of $\nu$s from Sun or Earth & \multirow{2}{*}{HK, DUNE, IceCube ...} \cr 
    & Boosted DM, slow-moving DM & Scattering, or up-scattering \& decay & \cr 
    \cline{2-4}
    & \multirow{2}{*}{Fixed target} &  Decay &  \multirow{2}{*}{DUNE, T2HK, SBN, FASER$\nu$, ...} \cr
    & & Scattering, or up-scattering \& decay & 
    \\
    \hline 
    \multirow{2}{*}{Milli-charged particles\cite{Batell:2022xau}} & Fixed target & \multirow{2}{*}{Scattering} & DUNE ND, T2HK ND, ... \cr
    & Atmosphere & & DUNE, HK, JUNO, ...
    \\ 
    \hline
    \hline 
    \end{tabular}
    }
    \caption{ Summary of the most significant experimental signatures for the BSM scenarios covered here. Example experiments sensitive to each scenario are also provided (see references for the full list). Abbreviations: Atm.=Atmospheric, $B$=Baryon number, CC=Charged Current, CE$\nu$NS=Coherent Elastic $\nu$-Nucleus Scattering, DM=Dark Matter,  FD=Far Detector, HE=High Energy, LED=Large Extra Dimensions, ND=Near Detector, NSI=Non-Standard Interactions, SN=Supernovae, UHE=Ultra-High Energy, DAR=Decay at rest. }
    \label{tab:my_label}
    \end{table}
    \end{landscape}
    \clearpage
}


\section{Introduction and Motivation}
\label{sec:introduction}

Since the discovery of neutrino oscillations, neutrino physics has been a field driven by data, and with the current and upcoming generation of neutrino experiments this will likely continue to be the case. Since the last Snowmass process, we have seen the field flourish in several areas simultaneously, from the large improvement of the precision at which we know neutrino oscillation parameters, with the first hints for CPV and mass ordering emerging from global fits, to the emergence of Coherent $\nu$-Nucleus Elastic Scattering (CE$\nu$NS) measurements as a completely new field of research, thanks to the recent developments in detector technology associated to dark matter direct detection experiments. We are also witnessing the blooming of neutrino astronomy, with more and better measurements at Icecube and KM3NeT, which have in turn triggered further initiatives to build neutrino telescopes worldwide. In parallel, new bounds on the neutrino mass from KATRIN and cosmological observations are also pushing down the upper bounds on the value of neutrino masses, and very recent measurements by KamLAND-Zen are already entering the region expected for neutrinoless double-beta decay if light neutrino masses follow an inverted ordering scheme.

On the oscillation front, while T2K and NO$\nu$A continue accumulating data and will provide very useful hints on the remaining neutrino oscillation parameters, new and more capable facilities will be needed to reach a high significance. These will require more powerful neutrino sources and larger, massive detectors, which unavoidably demands an excellent control of systematic uncertainties, given the large statistics expected.  
New facilities (such as DUNE, T2HK, and JUNO) are expected to improve the determination of neutrino mixing angles and mass splittings significantly. As we enter this precision era, one of the main goals of such experiments should be to thoroughly test and probe the three-neutrino paradigm, something which has been out of reach for the neutrino community thus far. 
In fact, while we are aware of the need to extend the Standard Model to account for non-zero neutrino masses, the possible phenomenological consequences of such extensions are endless. 
Beyond the Standard Model (BSM) effects in the neutrino sector may lead to significant alterations in the flavor composition and spectra of neutrinos produced across energy scales and in neutrino sources differently, and therefore an exploration on multiple fronts is crucial. For example, conventional neutrino beams offer the possibility to use a pulsed neutrino flux and to characterize the neutrino spectra with the aid of a near detector before any oscillations have taken place. On the other hand, the observation of neutrinos from supernovae would allow us to test our knowledge of neutrino interactions with matter in extremely dense environments and to probe neutrino self-interactions which otherwise would be inaccessible in the laboratory. Similarly, ultra-high energy astrophysical neutrinos reach our detectors after traveling extremely long distances and probe neutrino interactions in a very different energy regime. 

Due to the very elusive nature of the neutrino, neutrino experiments typically rely on massive detectors, a powerful source, or a combination of the two. The high-intensity proton beams that are required, in conjunction with powerful combinations of precision near detectors, provides opportunities to search for rare processes that were inconceivable in previous generations of neutrino experiments. The use of underground neutrino detectors enable searches of particles from cosmogenic sources or produced in the upper layers of the atmosphere, thanks to their sheer target mass. In addition, the improved capabilities of the detector will greatly enhance signal-to-background discrimination. This allows the use of neutrino experiments to probe weakly coupled sectors and has motivated the community to exploit them as \emph{diverse tools} to push the BSM physics in the Intensity Frontier forward as much as possible, complementing the phase space covered by Energy Frontier experiments. In this sense, the theoretical and phenomenological communities have embarked on a dedicated effort to design new strategies to probe models for dark sectors, which often result in stronger limits than dedicated direct detection or collider searches for dark matter. This is something that had already begun at the time of the last Snowmass process, and that has been widely expanded and pursued since then. Finally, leveraging the fundamental nature of these rare processes also broadens the opportunities for collaborations with the nuclear physics community, starting from precision measurements of neutrino-nucleus interaction cross-sections to utilization of rare-isotope facilities for dark sector particle searches.

This report summarizes the BSM opportunities that will become available at the next generation neutrino facilities, leveraging their capabilities. We present summaries of several topical areas of BSM physics, classified in two distinct categories: BSM physics opportunities in the neutrino sector (stemming from the existence of new neutrino interactions, and/or heavy neutrino states); and opportunities at neutrino facilities to explore and constrain BSM scenarios in other sectors (e.g. dark sectors). Each of these topics takes advantage of different aspects of future neutrino experiments, which are emphasized in each sub-topical whitepaper~\cite{Abdullahi:2022jlv,Arguelles:2022xxa,Dev:2022jbf,Abdullah:2022zue,Batell:2022xau,Berger:2022cab}. Our report concludes with an outlook of the exploration of BSM physics topics and a successful strategy for ensuring a continued pursuit of these goals.  


\section{Neutrino BSM Opportunities}
\label{sec:bsm-beams}



\subsection{Heavy Neutral Lepton Searches}\label{sec:hnl}

Heavy Neutral Leptons are right-handed neutrino partners to the 
Standard Model active neutrinos. Their existence
can provide elegant solutions to present open questions in fundamental
physics such as the origin of neutrino masses, the nature of dark matter and the observed matter antimatter asymmetry in the Universe.
These HNLs, named as such because they are significantly heavier than the Standard Model active neutrinos, are (quasi) sterile and are produced through mixing with the active neutrinos.
The allowed mass range for these putative particles is unknown and spans any value between a fraction of an eV up to the GUT scale. Hence HNLs are searched for with a large number of complementary experimental approaches, from oscillations and nuclear decays to energy frontier experiments. In this report, we refer to searches for HNL with masses above the eV scale, while for lighter states we refer the interested reader to the Snowmass whitepaper~\cite{Acero:2022wqg} and to the NF02 Topical Report~\cite{https://doi.org/10.48550/arxiv.2209.05352}. In the mass range between keV and TeV, a number of existing and new opportunities stand out for the hunt for HNLs in the coming decades, as outlined below. In addition to examining HNLs which only interact via neutrino mass mixing, we have also surveyed the phenomenological consequences of non-minimal HNLs which have additional interactions. The current and future status of HNL searches can be found in the recent Snowmass whitepaper~\cite{Abdullahi:2022jlv}. 

A novel approach to searching for HNLs involves exploiting energy-momentum conservation in nuclear reactions in which an electron neutrino or electron antineutrino is involved, such as beta decay and electron capture processes.
Proposed new experiments to probe these processes in the next decade will provide valuable handles to make a direct search for heavy neutral leptons in the full keV HNL mass range, with coupling sensitivities that will improve the present experimental reach by several orders of magnitude, in particular when including the envisaged potential upgrades of these experiments.

Present, future, and upgraded short-baseline experiments have a window to improve the sensitivity for HNL searches  in the mass region of 1-10 MeV.
The challenge for these experiments will be to have good handles on the background control, and ensure dedicated triggers for HNL decays in flight, which would allow coverage of a substantial extension of the present search region in that mass range. We strongly recommend the reactor experiment community to study and invest in this particular opportunity.

The prospects for discovering HNLs in the coming one or two decades in fixed-target experiment environments have been examined. This broad category includes many currently-operating and next-generation experiments, each with various approaches and physics goals (many of which are orthogonal to these beyond-the-Standard-Model searches). These can be  broadly categorized based on their experimental equipment deployed, and can be used to divide these  into searches from rare kaon decays, beam-dump setups, and searches in neutrino-beam environments. 
Complementarity among the different fixed-target probes is evident, but also when comparing with the other types of searches discussed e.g. for colliders as discussed below. Fixed-target searches offer some of the most promising sensitivity to discovering HNLs in the tens of MeV to few GeV range in the near future.

The LHC and possible future high energy colliders will offer excellent opportunities to search for heavy neutral leptons. With the full high-luminosity event statistics the CMS and ATLAS experiments can potentially reach values of couplings in the minimal HNL model on $|V_{e N}|^2$ and $|V_{\mu N}|^2$ down to or below $10^{-7}-10^{-8}$, in the mass region $m_N $ of 5-20 GeV. LHCb will extend the range to lower masses. One of the issues hindering the reach to smaller couplings at small mass hypotheses is the decreasing acceptance of the central detectors due to the  correspondingly increasing HNL lifetimes. 
Current proposals for new experiments at the LHC, made to overcome this limitation, are grouped in transverse and forward type of detectors. The transverse detector proposals encompass the MATHUSLA, CODEX-b, AL3X, ANIBUS and MAPP-LLP experiments. These are typically experiments optimized for searches for observing new weakly interacting neutral particle decays, and placed at distances of tens to more than a 
hundred meters away from the new particle production point.
Forward detectors, such as FASER, SND@LHC, the Forward Physics 
Facility and FACET, are located along the direction of the LHC beam line and are mostly sensitive to the production of new neutral particles 
originating in decays of mesons.

These additional detectors will cover an important part of the HNL 
parameter space, mostly for masses $m_N$ less than 5 GeV, and will be complementary to  experiments at high intensity fixed target experiments. The sensitivities  will be reaching values of $|V_{e N}|^2$ and $|V_{\mu N}|^2$ roughly down to $10^{-8}-10^{-9}$, possibly even lower values, in a mass region between 100 MeV and 5 GeV. 


In the more distant future, a facility like the FCC project could be realized. 
In terms of searches for HNLs, the FCC in its different complementary stages, can probe very large areas of the parameter space in the tens of GeV mass region, due to the copiously produced heavy bosons (Z's at the FCC-ee Z-factory and W's at a high luminosity FCC-hh hadron collider) that are large sources of neutrinos, and cover regions
that are not constrained by astrophysics or cosmology, and are complementary to beam dump and neutrino facilities. Heavy neutrinos with masses larger than 10 TeV can be searched for at the high energy frontier at the FCC-hh.

In the event of an HNL discovery, it will be important to study the HNL's (or HNLs') properties. This includes, but is not limited to, studying the mixing pattern(s) and flavor structure of the HNL(s) to determine if there is a connection to the observed light neutrino masses, as well as determining whether Lepton Number is conserved or violated, or equivalently, whether neutrinos and HNLs are Dirac or Majorana fermions. Either of these observations would revolutionize particle physics. In particular fixed target experiments would be in excellent position for such measurements if the HNLs happen to live in their covered parameter space, but present and future collider detectors will prepare for this too.

In addition to the terrestrial bounds discussed above, the Snowmass whitepaper~\cite{Abdullahi:2022jlv} also surveyed the landscape of constraints arising from solar, atmospheric, astrophysical, and cosmological considerations. The solar and atmospheric neutrino fluxes are large sources of naturally occurring neutrinos which can be utilized for HNL searches. These turn out to lend themselves particularly well to providing strong constraints on non-minimal HNLs such as those interacting with a transition magnetic moment. Likewise, the presence of HNLs in the early universe can potentially destroy the success of big bang nucleosynthesis.  These cosmological constraints provide complementary sensitivity to HNLs, reaching lower mixing angles than any existing terrestrial constraint. At the same time, to the extent that terrestrial and cosmological sensitivities overlap, there is the possibility of detecting HNLs which could require modifications to cosmology. 

For most of the future options and proposals given in Ref.~\cite{Abdullahi:2022jlv} -- both for the near and more distant future -- first estimates on the
sensitivity for HNL discoveries have been made, and demonstrate 
the potential HNL parameter space coverage.
Certainly further studies e.g on detector optimization are strongly desirable and needed for this important physics target. Such studies are encouraged to go beyond the simplest version of the HNL models, covering non-minimal scenarios.

We finish this section with the following suggestions in the interest of the community:

{\bf Suggestion 1:} We strongly suggest the experimental  program to pursue new ideas and make proposals for HNL sensitive experiments at the current existing accelerator facilities, as well as continue to explore the hunt for HNLs with  already existing detectors and/or upgrades. Some facilities were perhaps not designed for BSM particle hunt studies per se, but thanks to their high intensity proton source and the newly planned near detectors that will have excellent resolutions and efficiencies, these will become very competitive, and one should exploit this superb opportunity to ``upgrade'' the searches for such new particles to become a key part of these experiment's baseline physics program.

{\bf Suggestion 2:} It is suggested that future collider facilities take into account from the start the strong interest and need for searches for long lived particles in their infrastructure plans. Detector designs should from the start take searches for HNLs and LLPs in general as part of their main physics goals.

{\bf Suggestion 3:} In order to facilitate apples-to-apples comparisons, and for simplicity, we encourage experimental analyses to examine sensitivity to the electron-, muon-, and tau-HNL mixing angles separately. Of course if time allow,  other flavor assumptions are worth exploring, and possibly even more realistic.  
For example, a discussion of such new benchmarks is ongoing in the context of the FIPS workshop series~\cite{Agrawal:2021dbo}.

{\bf Suggestion 4:} The keV mass scale can be covered using nuclear processes, and the proposed experiments are very important. We should also make sure to capitalize on the existing and planned (upgraded) reactor experiments to cover the low MeV mass range.

{\bf Suggestion 5:} In the aftermath of HNL discovery, the most immediate question will be the experimental determination of HNL properties. One would like to extract the data-preferred mixing angles and HNL mass, determine the nature of the quantum statistical nature of the new particles and perhaps stress-test the assumption that the detection is consistent with minimal HNL couplings. We suggest experiments also consider how best to exploit the ability of their data to determine such observables in the case of a discovery.

\subsection{BSM Effects on Neutrino Flavor}
\label{sec:osci}

Neutrinos offer one of the most promising places to search for Beyond the Standard Model (BSM) effects. On the theoretical side their elusive nature, combined with their unknown mass mechanism, seems to indicate that the neutrino sector is indeed opening a window to new physics. On the experimental side, several long-standing anomalies have been reported in the past decades, providing a strong motivation to thoroughly test the standard three-neutrino oscillation paradigm. This can be done in three main ways. First, neutrino oscillations experiments are very precise interferometers, sensitive to subleading effects from new physics affecting neutrino flavor transitions. On a separate front, neutrino telescopes provide a unique avenue to probe BSM effects, given the very long distances traveled by the detected neutrinos (which range from the Earth radius to several gigaparsecs), as well as their ultra-high energies. Finally, astrophysical observations (such as a nearby core-collapse supernovae) in the upcoming decade will provide invaluable information and allow us to test neutrino propagation in extremely dense environments. The Snowmass white paper~\cite{Arguelles:2022xxa} summarizes the main BSM scenarios that can lead to new signals affecting neutrino flavor transitions and neutrino oscillations.

Until now neutrino physics has been driven by data, from the postulation of the neutrino by Pauli to the discovery of neutrino oscillations, which was awarded the Nobel Prize in 2015. While the upcoming generation of oscillation experiments aims to measure the leptonic CP phase and the neutrino mass ordering, it will also test the standard picture with an unprecedented level of precision. For the community to succeed in this goal, a rich experimental program is required, an effort which should be complemented with a similar one by the community working in phenomenology and theory. Here we highlight interesting and new opportunities to discover the presence of new physics affecting neutrino flavor transitions in the next two decades, as well as some of the challenges that will have to be faced in order to succeed in such a goal.

Testing neutrino oscillations in different environments is extremely useful to break degeneracies between standard and BSM parameters, and to make sure that the three-neutrino paradigm is robust. This entails contrasting the oscillation pattern in matter and vacuum,  as well as on different oscillations channels, and/or for experiments relying on different detection mechanisms, among others. In fact, since the last Snowmass process new data from T2K and NOvA have become available, enabling us to test the consistency of the standard neutrino oscillation picture. In the future, DUNE, JUNO and T2HK will continue this endeavor, with a much higher level of precision. This also extends to non-neutrino experiments: as an example, dark matter direct detection experiments are sensitive to BSM effects on the neutrino floor. With this goal in mind, it is worthwhile stressing the importance of global fits to neutrino data. In the past two decades these have proven to be extremely powerful to unveil subleading effects in the oscillation probabilities (providing evidence for a non-zero $\theta_{13}$ before it was determined experimentally); similarly, they will be critical to unveil new physics effects on neutrino flavor. 

Experimental collaborations are already working on joint fits for the determination of the neutrino mixing parameters (between T2K and NOvA, ongoing), and analogous efforts have been carried out by reactor experiments searching for eV-scale sterile neutrinos. It would be highly desirable to extend these to constrain other new physics models as well, to extract the most out of the available data. However, it is also important for collaborations to facilitate information to the public, so the whole community may analyze their data to look for BSM signals. Two notable examples in this context are the Icecube and COHERENT collaborations: their data release efforts have empowered the theoretical community to reinterpret their results and to derive new bounds on BSM scenarios, or to propose new experimental searches to unlock their full potential. In this effort the use of general (model-independent) parametrizations is critical, as it allows to easily recast the obtained experimental bounds to specific BSM models, and to extend the applicability of the analyses performed with the data. An example is the use of effective operators to include possible effects from new interactions at low energies, a similar approach also employed in other areas of particle physics (for example, in collider searches for BSM signals). Additional frameworks of special relevance in the neutrino sector include the parametrizations used to describe deviations from unitarity in the leptonic mixing matrix, or SME coefficients in scenarios where Lorentz symmetry is violated. Nevertheless, this needs to be complemented with an effort on the theory side, to make sure that the bounds on effective parameters are correctly interpreted, and matched onto viable models.

Finally, it is worthwhile stressing that, so far, new physics has been evading our main avenues to search for it, and thus it could yield unexpected signals. Because of this, experiments should be flexible and not single-purposed, and those experiments still in their designing stages should aim to expand their scope as much as possible. For example, at long-baseline oscillation experiments, strong emphasis is placed on the charged-current measurements for the standard oscillation program; 
however for BSM searches neutral current measurements are equally important. Also, experimental data may reveal new discoveries in new and unexpected ways. For example, at long-baseline oscillation experiments the neutrino flux measurements at the near detectors are assumed to be unoscillated so they can be used to extract the oscillation parameters at the far detector. However, near detectors themselves are sensitive to certain BSM effects; thus, unexpected anomalies (e.g. on the event rate normalization at near detectors) should not be completely dismissed. In fact, new generation neutrino oscillation facilities feature beams of unprecedented luminosity which, when observed by their near detectors, will provide a very powerful tool to explore new physics effects. However, given the very high statistics, the bottleneck for the sensitivity to such searches is often the level of understanding of the signal sample. That is, systematic uncertainties are critical and must be evaluated and modelled thoroughly in order to derive reliable constraints; a supporting program of flux and neutrino interaction measurements will also be needed. Similar restrictions apply to BSM searches using neutrino telescopes, which are subject to large uncertainties stemming from our poor knowledge of neutrino fluxes and cross sections at such high energies. At neutrino telescopes, improving neutrino flavor identification will be key in order to improve their capabilities to test for BSM effects.


\subsection{BSM Effects on Neutrino Scattering}

Neutrinos are a fundamental ingredient of the SM of particle physics and cosmology. Due to their
neutrality and weak interactions, however, their properties are the least understood among the SM particles. In particular, neutrinos can serve as a portal to BSM physics, imbuing them with new interactions. Remarkably, these interactions can be significantly larger than those provided by the electroweak (EW) gauge bosons, suggesting that multiple new observable phenomena may be enabled by such interactions. 

New neutrino interactions with SM matter fields would impact neutrino oscillations in presence of matter, as discussed for example in Ref.~\cite{Arguelles:2022xxa}. However, such interactions may also be tested directly in the laboratory, through precise measurements of neutrino scattering on electrons and nuclei. Among these efforts, the recent observation of Coherent Elastic neutrino-Nucleus Scattering (CE$\nu$NS) should be highlighted in the context of BSM searches. CE$\nu$NS is a process in which neutrinos
scatter coherently with a whole nucleus. Although the total cross section is large by neutrino standards, CE$\nu$NS has long proven difficult to detect, since the
deposited energy into the nucleus is $\sim$ keV. Since its first observation in 2017 by the COHERENT collaboration, the detection of CE$\nu$NS
has spawned a flurry of activities in high-energy physics, inspiring new constraints
on BSM physics, especially regarding the existence of new neutrino interactions since it provides constraints that are complementary to those derived from oscillation data. Interesting connections also exist between CE$\nu$NS and dark matter detection experimental programs: while solar CE$\nu$NS signals in future dark matter detectors constitute an important background to dark matter searches, its observation can also be used to extract bounds on new neutrino properties and interactions. 

The CE$\nu$NS process has important implications not only for high-energy physics, but also astrophysics, nuclear physics, and beyond. The main terrestrial sources of neutrinos used for CE$\nu$NS experiments include stopped-pion sources, nuclear reactors, $^{51}$Cr sources, and next-generation neutrino beams. Astrophysical sources of neutrinos can also be used to detect CE$\nu$Ns, including solar neutrinos, atmospheric neutrinos, and supernova neutrinos. A recent whitepaper~\cite{Abdullah:2022zue} discusses the scientific
importance of CE$\nu$NS, highlighting how present experiments such as COHERENT are
informing theory, and also how future experiments will provide a wealth of information
across the aforementioned fields of physics.

In the deep-inelastic regime, additional probes of new neutrino interactions are available: at high energies, BSM interactions may not only increase the size of the neutrino-nucleon cross section, but they may also affect the inelasticity distributions and provide significant deviations from the SM expectation. At energies above $\sim 10~\mathrm{TeV}$, the neutrino-nucleon cross section is measured by observing neutrino absorption in the Earth using data from neutrino telescopes. With current experiments such as Icecube, the total cross section has been measured up to $1-10~\mathrm{PeV}$. Future radio-detection experiments, as well as the Icecube Gen2 upgrade, will aim to extend the energy range up to $10^{20}~\mathrm{eV}$. Currently, inelasticity measurements have been obtained by Icecube using neutrino data up to 500~TeV; future radio-detection experiments may be able to extend this to substantially higher energies. However, it should also be stressed that in order to provide significant BSM constraints, the uncertainties in SM processes, including nuclear corrections and non-DIS processes, should be reduced with respect to present values.

Finally, while non-standard interactions with charged SM fermions are in some cases straightforwardly tested with traditional neutrino scattering and
oscillation experiments, it is remarkable that neutrino self-interactions, $\nu$SI, may also be probed through a variety of methods. Neutrino self-interactions arise in multitude of BSM scenarios, including models addressing neutrino mass generation or the production of dark
matter in the early Universe, as well as gauge extensions of the SM, among others. Below the scale of EW symmetry breaking, $\nu$SI can usually be described by a schematic interaction like $\nu \nu \phi$ where $\phi$ is a scalar or vector mediator particle. However, since neutrinos are part of an EW doublet, an
ultraviolet (UV) completion is required to consistently embed such an interaction within the SM. The related Snowmass whitepaper~\cite{Berryman:2022hds} focuses on $\nu$SI and discusses both the theoretical motivation and the physical systems in which they may lead to observable phenomena. 

Experimentally, neutrino self-interactions may be probed in cosmological and astrophysical environments as well as in the laboratory, providing important and complementary constraints across a broad range of parameter space: at relatively low scales (roughly between $\sim$eV and $\sim$MeV), $\nu$SI would impact cosmological observables including light element
abundances, the Cosmic Microwave Background (CMB), and the matter distribution in the Universe; for higher self-interaction scales up to $\mathcal{O}(100~\mathrm{MeV})$, supernovae and other astrophysical sources of neutrinos provide the strongest limits; finally, 
laboratory experiments can access the broadest range of self-interaction scales all the way up to
$\mathcal{O}(100~\mathrm{GeV})$. In order to thoroughly test the existence of $\nu$SI, it is imperative that the community considers all of these domains simultaneously. The whitepaper~\cite{Berryman:2022hds} demonstrates how several future experiments will further probe $\nu$SI, including theoretically well-motivated targets. 
\section{Non-Neutrino BSM Opportunities}
\label{sec:bsm-cosmos}

\subsection{Baryon Number Violation Searches}\label{sec:bnv}

The stability of ordinary matter has long been a subject of both theoretical and experimental interests.  The electron is stable because of electric charge conservation. On the other hand, the stability of the proton is guaranteed in the Standard Model by the accidental global symmetry of baryon number. In models of quark-lepton unification, such as the Grand Unified Theories (GUTs), baryon number is necessarily violated. As a result, the proton is no longer stable, and dominantly decays into $e^+\pi^0$ (for non-supersymmetric theories) or $K^+\bar{\nu}$ (for supersymmetric theories). This `smoking gun' prediction of GUTs which are otherwise inaccessible to laboratory experiments motivated the construction of large-scale water Cherenkov detectors like Kamiokande (later upgraded to Super-Kamiokande). Although there is no direct evidence of proton decay so far, but only stringent lower limits on the proton lifetime, it is important to continue the searches for proton decay and other baryon number violating (BNV) processes in general. In fact, the observed matter-antimatter asymmetry suggests that baryon number must be violated at some level. It is also important to keep in mind that the same experiments originally constructed to search for proton decay have now become truly multi-purpose experiments. In particular, they have played a major role in neutrino physics, starting with the serendipitous detection of SN1987A neutrinos, as well as  the discovery of neutrino oscillations. 

Therefore, the significance of current and next-generation neutrino experiments {\it simultaneously} searching for baryon number violation and studying neutrino properties cannot be overemphasized. While the main focus of the BNV experiments is on proton decay searches, there also exist other equally important baryon and/or lepton number violating processes, such as dinucleon decays and neutron-antineutron oscillations which must be studied as well along with their experimental detection prospects. Possible connections of BNV observables to other beyond the Standard Model physics, such as neutrino mass, baryogenesis, dark matter, flavor physics, and gravitational waves are also being explored. The recent lattice developments for the relevant nucleon and nuclear matrix elements of effective BNV operators are also crucial for reducing the theoretical uncertainties in the BNV predictions.  


Experiments have long sought evidence of the decay of the proton as proof of physics beyond the Standard Model.  The lower limit on the proton's lifetime is currently of order $10^{34}$~years.  Experimental searches that seek to probe beyond this limit therefore need a huge source of protons and years of exposure.
The large mass and long operation times of detectors used for observation of neutrino oscillations make them well-suited for searches for baryon number violation, including nucleon decay and neutron-antineutron oscillations.  The Super-Kamiokande neutrino experiment, which uses a 22.5~kton water Cherenkov detector and has been in operation since 1996, has published leading limits on 30 baryon number violating processes. Next generation neutrino detectors, such as DUNE (40~kton liquid argon TPC), Hyper-Kamiokande (190~kton water Cherenkov), and JUNO (20~kton liquid scintillator), all include baryon number violation searches as a major component of their physics programs and hope to improve upon the limits set by Super-Kamiokande, if not observe baryon number violation for the first time.  




Detector mass is a crucial characteristic in next-generation baryon number violation searches.  For small detectors, the exposure required to improve upon limits already set by Super-Kamiokande can exceed the likely lifetime of the experiment. Clearly Hyper-Kamiokande has the advantage in that respect. That being said, detector technology is also extremely important; DUNE's excellent imaging capabilities and JUNO's superb timing resolution offer advantages in some channels over Hyper-Kamiokande's larger mass.  NOvA, a currently-running neutrino experiment with a 14~kton segmented liquid scintillator detector, is developing a search for neutron-antineutron oscillations that could potentially have sensitivity comparable to current limits. Theia is a proposed water-based liquid scintillator detector that would combine the advantages of the large mass of a water Cherenkov detector with the good resolution of a liquid scintillator detector.  With this worldwide program, should a baryon number violation signal be observed by any one detector in the next generation, confirmation from other detectors using different technologies would provide powerful evidence of physics beyond the Standard Model.

In addition to detector mass and technology, simulation and analysis techniques can also affect the potential of these searches.  As with neutrino interactions, the experimental community has come to understand how important nuclear effects are in predicting the characteristics of final state particles.  Final state interactions in the nucleus alter the multiplicity and momenta of final state particles.  Uncertainties in modeling final state interactions therefore introduce uncertainties into the signal efficiency estimates and lifetime limits.  Furthermore, analysis techniques are continually improving.  For example, Super-Kamiokande made improvements to the search for proton decay via $p \to e^+\pi^0$ by reducing backgrounds via neutron tagging.  Potential improvements to searches in a liquid argon TPC could come from tagging of nuclear de-excitations.

The experimental neutrino physics community has long been conducting searches for baryon number violation using neutrino detectors.  The next generation of neutrino detectors will allow the continued pursuit of this goal, with massive detectors and continually improving analysis techniques~\cite{Dev:2022jbf}.

\subsection{Cosmogenic Dark Matter and Exotic Particle Searches}\label{sec:dm-cosmos}
Signals from outer space and their detection have been playing an important role in particle physics, especially in discoveries of and searches for physics beyond the Standard Model (BSM); beyond the evidence of dark matter (DM), for example, the neutrinos produced from dark matter annihilation are important for indirect DM searches. 
Moreover, a wide range of new, well-motivated physics models and dark-sector scenarios have been proposed in the last decade, predicting cosmogenic signals complementary to those in the conventional direct detection of particle-like dark matter.
Most notably, various mechanisms to produce (semi-)relativistic DM particles in the present universe (e.g., boosted dark matter) have been put forward, while being consistent with current observational and experimental constraints on DM. 
The resulting signals often have less intense and more energetic fluxes, to which underground, kiloton-scale neutrino detectors can be readily sensitive.
In addition, the scattering of slow-moving DM can give rise to a sizable energy deposit if the underlying dark-sector model allows for a large mass difference between the initial and final state particles,
and neutrino detectors are excellent places to explore these opportunities. 

Detectors based on different technologies are complementary for probing diverse models and scenarios.
For example, water Cherenkov detectors normally have large mass, nanosecond-level time resolution, and MeV-level detection thresholds for electrons, leading to the most stringent limits as of today on boosted dark matter originating from the Galactic Center and  Sun~\cite{Agashe:2014yua, Berger:2014sqa, Kong:2014mia, Super-Kamiokande:2017dch} as well as on ``dark cosmic rays'' of DM accelerated in astrophysical sources.
While long-string water Cherenkov detectors are uniquely suitable for TeV-scale signals, liquid-argon time-projection chambers (LArTPCs) may have an advantage searching for hadronic boosted DM interactions, owing to their moderately large nuclei and the capability of detecting protons with kinetic energy down to a few tens of MeV.
As various neutrino experiments are currently collecting data or will be operational in the future, a vast swath of parameter space will soon be explored.

Spanning over a wide energy range, the cosmogenic BSM searches broaden the physics cases at neutrino detectors and enhance the research and development of experimental techniques and analysis strategies.
Neutrinos from natural sources, such as solar, atmospheric, and astrophysical neutrinos, contribute as the main background, and understanding of such neutrino fluxes and interaction cross sections is crucial.
Searches for MeV-scale signals also encounter the background sources from radiological materials, which make recording such signals in real time challenging.
Innovative development on detectors, triggering systems, reconstruction algorithms, etc. will be helpful to comprehensively collect and analyze physics activities of interest.

A recent white paper~\cite{Berger:2022cab} is devoted to discussing the scientific importance of cosmogenic dark matter and exotic particle searches, not only overviewing the recent efforts in both the theory and the experiment communities but also providing future perspectives and directions on this research branch. 
A landscape of technologies used in neutrino detectors and their complementarity is discussed, and the current and developing analysis strategies are outlined.

\subsection{Beam-Originating Dark Sector Particle Searches}\label{sec:beam-dm}

The idea of a dark sector (also referred to as a hidden sector or secluded sector) implies the existence of new states that are not charged under the known strong, weak, and electromagnetic forces, but that are weakly coupled to the Standard Model via a mediator portal interaction. This concept is well motivated from a variety of perspectives as dark sectors may provide novel answers to a some of the big open questions in particles physics, including dark matter and neutrino masses, among others. The portal concept, based on straightforward effective field theory reasoning, provides a systematic framework for the theoretical and experimental investigation of dark sectors. Furthermore, a number of creative dark sector models have been proposed to explain a variety experimental anomalies.

An expansive experimental program is emerging to explore the dark sector, and neutrino beam experiments have a critical role to play in these investigations. Modern accelerator-based neutrino experiments feature enormous proton beam-target collision luminosities, which can supply copious secondary forward fluxes of dark sector particles. Contemporary neutrino detectors benefit from large active masses and volumes, excellent particle identification and reconstruction capabilities, and capacities for precision energy, spatial, and timing measurements, which can be leveraged to detect a variety of rare dark sector signals and distinguish them from beam-related and/or cosmic backgrounds. These experiments are particularly well suited to the study of hadrophilic and neutrinophilic dark sector interactions. The coming decade and beyond promises to be an exciting era for dark sector research, with a number of neutrino beam experiments already in operation and several ambitious planned projects on the horizon. 

The past decade has witnessed intense theoretical exploration of dark sector models and their phenomenology, including the novel signatures and promising search prospects at neutrino beam experiments. A recent whitepaper~\cite{Batell:2022xau} reviews the status of a broad range of dark sector models, including scenarios featuring the vector portal, Higgs portal, neutrino portal, axion-like-particle (ALP) portal, dark neutrinos, and neutrino-philic interactions.  Collectively, these theoretical scenarios motivate a broad suite of searches, including  long-lived-particle decays to a variety of visible final states, elastic and/or inelastic scattering with detector electrons or nuclei, neutrino up-scattering to dark neutrinos followed by visible decays, and modifications to neutrino scattering processes. Current and future neutrino beam experiments will be able to probe large regions of uncharted parameter space. 

A effective and robust dark sector search program necessitates accurate simulation tools for both the myriad dark sector particle production channels and the rich array of detectable signatures. Several challenges must be met in the development of these tools, including modeling the complex target geometry and horn, an accounting of nuclear physics effects in production and detection,  the capability for fast detector simulation, and the identification/reconstruction of unique signal topologies, to name a few. 
So far, phenomenological studies have utilized a hodgepodge of publicly available event generators in tandem with home-grown codes to simulate signals and backgrounds, design mock analyses, and derive sensitivity estimates. 
However, in most cases this approach is not suitable for experimental analyses, and the development of packages that can be readily integrated into the existing simulation frameworks used by the collaborations is one key direction that calls for immediate effort. Furthermore, the development of novel reconstruction and analyses methods, perhaps including the use of modern machine learning methods, is another important arena where improvement can be anticipated. 

An important aspect in the search for dark sector particles from beam interactions is precise understanding of the backgrounds from neutrino-nucleus interactions.  For this, improving the neutrino-nucleus interaction model based on experimental measurements and reflecting them into the simulations tools are essential elements.  To accomplish this, an effective way of collaborating with the nuclear physics community must be sought and implemented in a timely fashion to strengthen our understanding of backgrounds.
Finally, the powerful and precision near detector complex is critical for leveraging the full beam power at the future neutrino experiments, given that the dark sector particles of interest are resulting from beam interactions in the neutrino target.  

Neutrino beam experiments represent an important front in the quest to explore the dark sector. Experiments currently in operation and those coming online over the next decade hold the promise to significantly advance these studies, yet there is still much important work to be done to realize their full physics potential, perhaps most notably in the development of robust and versatile simulation tools. Given the exciting array of opportunities outlined in this section, dark sector searches will form an exciting and even vital part of the broader physics program at existing and future neutrino beam experiments.



\section{Diversity, Equity, and Inclusion}
\label{sec:DEI}

As we have engaged with the broader community through the Snowmass process, we have taken steps to ensure that a wide diversity of contributions are included. It is crucial that we build a diverse, equitable, and maximally inclusive community of physicists across disparate domains. This goal has been reflected in the authorship and editorship of the white papers within NF03's purview.

\section{Summary}
\label{sec:conclusions}

This report has surveyed some of the high-priority BSM opportunities which can be explored at upcoming neutrino experiments. For a quick guide to some of these, we refer the reader to Table~\ref{tab:my_label}. 

To maximize coverage of the potential afforded by near-term experimental neutrino program, we reiterate a few key points: (1) a variety of BSM physics can be powerfully explored at neutrino facilities; (2) in order to extract their full potential, BSM searches need to be included among the main experimental goals since this may have an impact on the experimental design of future facilities;
(3) it would be desirable for collaborations to derive BSM constraints in a way that is as model independent as possible; (4) there is a clear and urgent need for improved simulation tools capable of producing BSM signals and precisely predicting interactions of their backgrounds, including that of neutrinos; and (5) finally, for the community to extract the most from facilities and their concomitant data, a broader and closer collaboration with other particle physics frontiers and the nuclear physics community would be required.  

\section{Acknowledgements}
\label{sec:acknowledgements}

We would like to warmly thank Phil Barbeau, Brian Batell, Albert de Roeck, Bhupal Dev, David Vanegas Forero, Teppei Katori, Doojin Kim, Raimund Strauss, Louis Strigari, and Yun-Tse Tsai for their work as editors of contributed Snowmass white papers. We would also like to thank all the authors who contributed to Snowmass NF03 white papers, and the members of the community who provided comments and feedback on the present report.




\renewcommand{\refname}{References}


\bibliographystyle{utphys}

\bibliography{common/tdr-citedb}

\end{document}